\documentclass[letterpaper,11pt]{article}

\usepackage{amssymb}
\usepackage{graphicx}
\usepackage{mathptmx}
\usepackage{epstopdf}
\usepackage{hyperref}

\newcommand{\beq}{\begin{equation}}
\newcommand{\eeq}{\end{equation}}
\newcommand{\beqa}{\begin{eqnarray}}
\newcommand{\eeqa}{\end{eqnarray}}

\title{Solar X-ray Flare Hazards on the Surface of Mars\footnote{To be
published in \emph{Planetary and Space Science}.}}

\author{David S. Smith\footnote{Corresponding author} ~and John Scalo\footnote{E-mail:
\texttt{\{dss,parrot\}@astro.as.utexas.edu}}}

\date{Department of Astronomy, The University of Texas at
Austin, Austin, TX 78712}

\begin{document}

\maketitle

\begin{abstract}

{\normalsize Putative organisms on the Martian surface would be exposed to potentially high doses of ionizing radiation during strong solar X-ray flares.
We extrapolate the observed flare frequency-energy release scaling
relation to total X-ray energies much larger than seen so far for the
sun, an assumption supported by observations of flares on other solar-
and subsolar-mass main sequence stars.  Flare spectra are taken as
power laws, with the logarithmic slope a parameter based on the observed
statistics of the most energetic hard X-ray flare spectra.  We calculate
the surficial reprocessed spectra using a Monte Carlo code we developed
for the transport of X-rays and gamma rays, including photoabsorption
and detailed Compton scattering.  Biological doses from indirect
genome damage are calculated for each parameterized flare spectrum by
integration over the X-ray opacity of water.  The resulting doses depend
sensitively on spectral slope, which varies greatly and unsystematically
for solar flares.  Using the roughly uniform observed distribution of
spectral slopes, we estimate the mean waiting time for solar flares
producing a given biological dose of ionizing radiation on Mars and
compare with lethal dose data for a wide range of terrestrial organisms.
These timescales range from decades for significant human health risk to
0.5 Myr for \emph{D. radiodurans} lethality. Such doses require total
flare energies of $10^{33}$--$10^{38}$ erg, the lower range of which
has been observed for other stars.  Flares are intermittent bursts,
so acute lethality will only occur on the sunward hemisphere during a
sufficiently energetic flare, unlike low-dose-rate, extended damage by
cosmic rays.  We estimate the soil and CO$_2$ ice columns required to
provide $1/e$ shielding as 4--9 g cm$^{-2}$, depending on flare mean
energy and atmospheric column density. Topographic altitude variations
give a factor of two variation in dose for a given flare.  Life in ice
layers that may exist $\sim 100$ g cm$^{-2}$ below the surface would
be well protected.  Finally, we point out that designing spacesuits to
sufficiently block this radiation on Mars missions may be difficult,
given the conflict between solutions for lightweight protection from
energetic particles and those from X-rays.}

\end{abstract}

\section{Introduction}\label{sec:intro}

Habitability for all types of life---from human to microbial---is a central
problem for the next decades of Mars exploration.  In particular,
knowledge of surficial radiation doses from all possible sources is
essential.  Cordoba-Jabonero et al. (2003) have calculated in detail the
transfer of ultraviolet (UV) radiation in the Martian atmosphere and
folded the surface flux with a DNA action spectrum to obtain UV dose
rates (see also Cockell et al., 2000, Mancinelli and Klovstad, 2000).
Schuerger et al. (2003) treated the same problem experimentally,
with the survival of spores of \emph{Bacillus subtilis} in a Mars
simulation chamber.  Molina-Cuberos et al. (2001), Pavlov et al. (2002),
and De Angelis et al. (2004) have estimated the fluxes of Galactic cosmic-ray
particles on the present and ancient Martian surface and subsurface.
Pavlov et al. find that the maximum dose rate due to Galactic cosmic
rays is about 0.2 Gray yr$^{-1}$, using dosimetry units in which 1 Gray
(Gy) $\equiv$ 100 rad and 1 rad $\equiv$ 100 erg g$^{- 1}$ absorbed.
This dose rate occurs at a depth of about 25 g cm$^{-2}$.  Measurements
and detailed simulations of the cosmic-ray dose rate have been carried
out by the Martian Radiative Environment Experiment (MARIE) on the Mars
Odyssey Orbiter (see Saganti et al., 2004), with measured and modeled
surface values mostly around 0.06 Gy yr$^{-1}$.  Integrated over a
microorganism's lifetime, the corresponding doses are smaller than the
lethal dose for any terrestrial microorganism, but the whole-body human
dose would be significant over decades, suggesting problems for manned
Mars missions. The dose due to ionizing radiation is also of interest
concerning natural transfer of microorganisms between solar system bodies,
in particular Mars and Earth (Mileikowsky et al., 2000).

X-rays from the most energetic solar flares are potentially important for
Martian habitability,
but have not yet been treated. The steady solar coronal X-ray, EUV, and
FUV emission is currently fairly small (see G\"udel et al., 2003). In
contrast, during a solar flare, the flux of ionizing photons can increase
by orders of magnitude, and large dose rates are possible. Unlike on
Earth, the Martian atmosphere is currently thin enough that a significant
fraction of this radiation can arrive at the surface, where it may lead to
genetic damage and mutations in terrestrial-like organisms.  The fraction
of energy absorbed in the atmosphere or reprocessed to UV (see Smith
et al., 2004a,b) is minimal.  Moreover, the X-ray flux from
the flares of interest here are orders of magnitude larger than the Martian keV
X-ray fluxes observed by the Chandra X-ray Observatory (Dennerl, 2002)
that are probably generated in the Martian atmosphere by solar X-rays
(see Cravens and Maurelis, 2001) and the solar wind (see Gunell et
al., 2004)

Evolution of enhanced resistance to the X-ray flares discussed in the
present work does not seem possible for anything analogous to terrestrial
evolution. The flare durations are very short (about an hour), and the
times between exposures are orders of magnitude larger than organism
lifetimes, so no adaptation could occur.  Thus highly intermittent flare
radiation can be lethal in a way that gradually increasing UV (or nearly
constant cosmic rays) cannot.

In this paper, we focus on the following question: At the Martian
surface, for a strong solar X-ray flare of given total energy and
spectrum, how does the biological dose compare to the lethal doses
of various terrestrial organisms?  We then extrapolate the recurrence
frequency-energy release relation for solar flares to estimate the mean
time between such lethal events for a variety of organisms.  Finally,
we contrast X-ray dose rates with that of cosmic rays and discuss
shielding considerations.

\section{Solar Flare Properties}\label{sec:flareprops}

\subsection{Spectra}\label{sec:fp_spectra}

X-rays and $\gamma$-rays with energies between 10$^{-1}$ and 10$^6$
keV\footnote{1 \AA $=$ 12.4 keV} comprise a significant fraction of
solar flare photon emission (e.g. Haisch et al., 1991; Hudson, 1991;
Kanbach et al., 1993; Ryan, 2000).  Hard X-ray (hereafter HXR) spectra
are often taken to peak around 10--25 keV (Crosby et al., 1993; but see
Krucker and Lin, 2002), but this is very uncertain because it is near
the sensitivity cutoff of most pre-RHESSI instruments (see Battaglia
et al., 2005).  Flare X-ray spectra are often fit by a single power
law (Crosby et al., 1993; Bromund et al., 1995; Veronig et al., 2002)
or two piecewise power laws (Krucker and Lin, 2002).  The range of
the estimated power-law spectral index (log-log slope) $p$ is large
for hard X-ray flares (see Lee et al. 1993, Bromund et al. 1995; also
Qiu et al. 2004 for HXR microflares): most flares have a $p$ between
2.5 and 6, with a median around 4 and little correlation with total
X-ray output.  Using results from RHESSI, Battaglia et al. (2005) find
that the non-thermal (hard X-ray) emission spectral index is correlated
with total energy release, such that the larger flares have flatter
(harder) spectra.   These were relatively low-energy flares, but if
a similar relation holds for the most powerful flares, then the large
solar X-ray flares described here may deliver even larger doses than
we estimate (see \S\ref{sec:doses}).  Given the uncertainty described,
we take the slope of the incident photon spectrum as a parameter (see
\S\ref{sec:radtran}), constrained by the observed distribution of $p$,
and assume no correlation between spectral index and total energy output.
As mentioned above, this likely gives a lower limit on the dose estimates.

\subsection{Durations}\label{sec:fp_dur}

Solar flares have a complex temporal structure.  The distribution of flare
durations has been estimated using several datasets and characteristically
fits a decreasing power law (e.g. Lee et al., 1993).  Most HXR flares
have short durations ($\lesssim$ 20 sec), with a loose correlation
between duration and total energy release.  We are only concerned with the
rare, highest-fluence flares, however, with durations found by Crosby et
al. (1993) for HXR and Veronig et al. (2002) for SXR flares of about 15
min, with the more energetic flares tending to last longer.  The X-class
data of Veronig et al. have a median duration three times longer than the
B-class data (30 min and 10 min, respectively).  Just as with spectral
slope, the flare-to-flare variation in duration is large and difficult
to characterize quantitatively.  For example, some flares emit much of
their energy as gamma rays over several hours (see Ryan, 2000).

For our purposes, though, the duration is unimportant; the acute lethal
dose is almost always independent of dose rate (Sparrow et al., 1967),
and thus depends on fluence, not flux. Note that almost all flares have
durations much smaller than a Martian day, suggesting that mutation and
lethality could occur only on one hemisphere at a time. On the other
hand, flare events clustered over many days are sometimes observed (e.g.,
the Halloween flares of Oct--Nov 2003; Woods et al., 2004) that could
progressively irradiate the whole planet.

\subsection{Energy Release}\label{sec:fp_er}

The total photon energy release in flares is difficult to estimate and
varies by at least 10$^8$ from flare to flare, but a large number of
studies using EUV, soft X-ray (SXR), or hard X-ray (HXR) satellite events
roughly agree that the differential distribution of flare energy releases
$dN/dW$ is a power law with index about $-1.6$ to $-1.8$ over at least
six orders of magnitude in total energy release $W$ (see Hudson, 1991;
Lee et al., 1993; Crosby et al., 1993; Bromund et al., 1995; Aschwanden et
al., 2000; Lin et al., 2001; G\"udel et al., 2003; Qiu et al., 2004; and
references therein), although a somewhat steeper index ($-2.0$) has been
inferred from a very large sample of SXR flares (Veronig et al., 2002).

Eleven large X-class flares occurred during the extraordinary solar
outbursts between 18 October 2003 and 5 November 2003, with SXR
releases in the GOES 1--8 \AA\,($\sim$ 2--10 keV) band peaking at about
$2\times10^{31}$ erg for an effective duration of 30 min.  Observations
using the SORCE instrument's Total Irradiance Monitor yielded a total
flare energy at all wavelengths for the somewhat weaker 28 October flare
of $4.6\times10^{32}$ erg (Woods et al., 2004).  Radiation and charged
particles from these flares compressed the Earth's Van Allen belt to
within 20,000 km of the surface (Baker et al., 2004), temporarily damaged
the orbiting Mars Odyssey communication instruments, and significantly
reduced stratospheric ozone levels (Randall et al., 2005).

Based on astronomical observations, we believe it reasonable to infer
that much more energetic solar flares have occurred in the past:

1. There is little doubt that solar-like stars can produce very
energetic, and frequent, X-ray flares.   Data presented by Audard et
al. (2000) indicate SXR flares (0.01--10 keV) of energy greater than
3--5 $\times10^{34}$ erg for the young, solar-like stars 47 Cas and
EK Dra (age $\sim$ 100 Myr) and $2\times10^{33}$ erg for an older,
solar-like star $\kappa$ Cet (age $\sim$ 1 Gyr), with frequencies of
about \emph{one per ten days} at these energies.  Considering the short
sampling period and the form of the derived frequency-energy release
relation, more energetic flares seem likely, though less frequent.

2. Schaefer et al. (2000) have identified nine ``superflares'' with
energy outputs of $10^{33}$ to $10^{38}$ erg on otherwise normal F8--G8
main sequence stars. Two of these stars were solar-like and produced
X-ray flares with energies about 100--1000 times larger than the largest
observed solar X-ray flare. These flares cannot be attributed to binaries,
rapid rotation, or youth, and may in fact be common in solar-type stars,
although cannabalization of giant planets has been suggested (Rubenstein and
Schaefer, 2000).
 
3. Stothers (1980) interpreted  NO$_3^-$ abundance spikes in
several Antarctic ice cores as due to flares with energies around
$10^{32}$--$10^{33}$ erg with a recurrence timescale of $\sim$ 10$^2$ yr,
consistent with the extrapolation of the frequency-energy relation we
adopt below in Eq.~\ref{eq:meantime}.

4. On lower mass dMe main-sequence stars, flares are more frequent,
with photon releases exceeding 10$^{34}$--10$^{35}$ erg observed,
often serendipitously (Hawley and Pettersen, 1991; Pagano et al., 1997;
Liebert et al., 1999; Favata et al., 2000; Christian et al., 2003).
These energies are enormous, considering that the bolometric luminosities
of these stars are much smaller than that of the sun.

Although flares of such large energy release have not been observed in
the sun, these data for other stars strongly suggest that the maximum
observed is limited only by duration of observation.  We have found no
compelling physical argument for an upper limit to flare energy releases
below the levels discussed in this paper, but the slope of the empirical
frequency-energy release relation suggests that some upper limit must
exist to avoid energy divergence (Hudson, 1991).  Limits on energetic
\emph{particle} fluxes (e.g., Lingenfelter and Hudson, 1980) are due to
self-confinement effects and cannot be applied to photons.  Given the above
considerations, we expect the the maximum flare release to be \emph{at
least} $10^{34}$ erg and assume this to be true in what follows.

\section{Radiative Transfer Method}\label{sec:radtran}

We calculate in detail the transfer of the incident ionizing
radiation through the Martian atmosphere using a previously
developed Monte Carlo code (Smith et al., 2004b) that accurately
treats Compton scattering and X-ray photoabsorption.  We use
the empirical approximation of Setlow and Pollard (1962) for the
photoabsorption cross section for energies greater than the K edge:
\beq \sigma_\mathrm{pa} = 2.04\times10^{-30}(1+0.008Z)\left({Z\over
E}\right)^3 \;\mathrm{cm}^2,\label{eq:sigma_pa}\eeq where $E$ is the
energy in units of 511 keV and $Z$ is the atomic number of the target
atom.  We have compared this with the NIST database of cross sections and
find good agreement.  Compton scattering cross sections are calculated
using the full Klein-Nishina formula.  We assume that the atmospheric
gas density falls off exponentially with height, with a scale height of
11 km and a total column density of 16 or 22 g cm$^{-2}$, corresponding
to rough limits set by the freezing and thawing of the polar ice caps.
This model works well because the high-energy radiative processes involved
here are relatively insensitive to the exact vertical distribution and
composition of gas, instead depending mainly on the total column density
and mean molecular weight.

Solar flare photon number spectra are assumed to be distributed
as $E^{-p}$, with $2\le p\le 6$ (see \S\ref{sec:flareprops}).  For
calculational purposes, the flare spectrum is assumed to extend from
10 keV to 511 keV.  The 10-keV lower limit is taken because photons
below this energy are nearly completely absorbed by the atmosphere
and unimportant to the surficial dose, while the upper limit is set
high enough that a negligible number of incident photons are at higher
energies, even for the shallowest spectra (lowest $p$).  Given a lower
limit $E_\mathrm{min}$ and a specified normalization $N_\mathrm{tot}$,
the differential number spectrum can be written (neglecting any upper
cutoff for simplicity) \beq {dN\over dE} = (p-1)\; N_\mathrm{tot}\;
E_\mathrm{min}^{p-1}\; E^{-p},\eeq and the average energy in the
spectrum is obtained by integrating between the minimum and maximum
cutoff energies: \beq \langle E \rangle = {1\over N_\mathrm{tot}}
\int_{E_\mathrm{min}}^{E_\mathrm{max}} E {dN\over dE}\; dE.\eeq This is
the spectrum that is propagated through the atmosphere using the Monte
Carlo code. Note that henceforth the total energy release figures refer
only to the energy between 10 and 511 keV.

After transport through the atmosphere, the photon fluence is converted to
a biological dose by assuming that the surficial radiation is absorbed
by pure water.  Much of the damage by ionizing radiation is thought
to be ``indirect,'' involving chemical reactions initiated by energy
deposited in the bulk cell water or first hydration layer rather than
``direct'' ionization of DNA (von Sonntag, 1987; see Ward, 1999, and
references therein), although this terminology is now recognized as an
oversimplification (see Fielden and O'Neill, 1991).  We then estimate
the dose by integrating the surficial energy spectrum $E\; dN/dE$
over the energy-dependent opacity of water $\kappa_\mathrm{w}(E)$:
\beq D = \int_{E_\mathrm{min}}^{E_\mathrm{max}} \kappa_\mathrm{w}(E)\;
E\, {dN\over dE}\; dE.\eeq The water opacity in our code includes both
Compton scattering and photoabsorption. It should also be noted that the
biological quality factor and relative biological effectiveness (RBE)
are both near unity for X-rays and $\gamma$-rays, so no adjustment is
required to compare to empirical lethal doses, as in cosmic-ray studies.

Most of the uncertainty in our calculations is due to the fact that the
total X-ray opacity of the Martian atmosphere depends sensitively on the
uncertain energy spectrum of the flare photons, since the photoabsorption
optical depth varies approximately as $\exp(-E^{-3})$.  Given a particular
flare spectrum, our results are accurate, but the spectral slopes vary
greatly from flare to flare, even given the same total energy release (see
\S\ref{sec:flareprops}).  Thus we use the slope, $p$, as a free parameter
that varies between 2 and 6 and present the results for a range in $p$.
Later we integrate over an approximate representation of the observed
frequency distribution of $p$ for hard X-ray solar flares to derive
appropriate average values for waiting times as a function of dose.

\section{Results}\label{sec:results}

\subsection{Calculated Surficial X-Ray Spectra}\label{sec:spectra}

\begin{figure}
  \centering
  \includegraphics[height=0.6\textheight,angle=270]{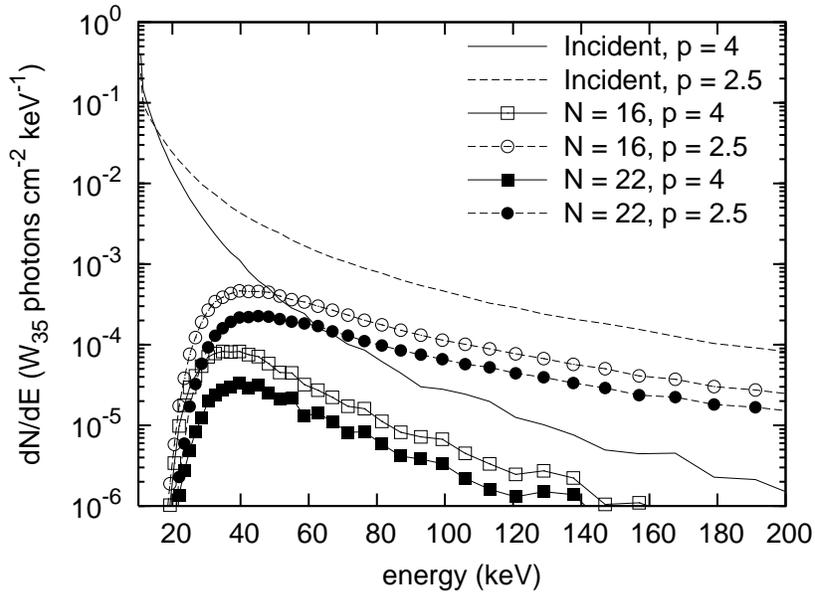}
\caption{Surficial, reprocessed, X-ray flare spectra, assuming a
parameterized, power-law model spectrum (see \S\ref{sec:radtran}).
Two incident spectra are shown, with power-law spectral indices of $p=2.5$
and $p=4$.  For each of these spectra, associated surficial spectra
are shown for two different column densities: 16 and 22 g cm$^{-2}$.
Attenuation due to photoabsorption is most severe at low energies, leading
to a sharp decline below about 20 keV in the flux per unit energy.  Also,
the original incident spectrum falls off at high energy, so the combined
effect leads to a peak at intermediate energies (30--50 keV here). }
  \label{fig:spectra}
\end{figure}

The Martian atmosphere, though thin by terrestrial standards,
significantly attenuates the incident flare radiation, but the high
fluences associated with the most massive flares lead to extreme
irradiation of the surface nonetheless. We show in Fig. \ref{fig:spectra}
the surficial fluence spectra from a $10^{35}$ erg flare for two
atmospheric column densities ($\Sigma=16$ and $\Sigma=22$ g cm$^{-2}$)
and two spectral indices ($p=4$ and $p=2.5$), together with the
above-atmosphere, incident spectra. The vertical axis scales linearly
with flare energy release (we have used 10$^{35}$ erg for this example).
In general, for all column densities, spectra with lower mean energies
(higher $p$) are attenuated more because of the rapidly rising ($E^{-3}$)
photoabsorption opacity.  This effect, combined with the falloff of the
incident spectrum at high energies, together creates a peak in the
surficial spectrum at moderate energies (30--50 keV here).  Additionally,
increasing the atmospheric column increases the attenuation at all
energies, as one would expect.

\subsection{Estimated Biological Doses}\label{sec:doses}

Ionizing radiation alters DNA through direct ionization and indirect
chemical interactions involving diffusion of radical products of
ionization and dissociation in the cytoplasm.  Resulting biochemical effects
include single- and double-strand breaks, base damage or abasic sites,
multiply damaged sites, oxidized base clusters, and cross-linking within
DNA or with proteins (see Becker and Sevilla, 1993; Ward, 1999). An
amazing suite of repair processes and associated enzymes repair the
damage (Friedberg, 2003, and references therein), but generally for
each organism, above some critical amount of absorbed energy, fecundity
greatly diminishes.  Biological damage is quantified in terms of the
amount of energy absorbed, or dose, typically in units of rads or Grays
(1 rad $\equiv$ 100 erg g$^{-1}$ absorbed; 1 Gray $\equiv$ 100 rad).
The lethal dose is often defined as the dose that kills a particular fraction of
a laboratory population.  The dose for reducing the population by $1/e$
is denoted $D_{37}$, while the dose for reduction of the population by
90\% is $D_{90}$.

\begin{figure}
  \centering
  \includegraphics[height=0.6\textheight,angle=270]{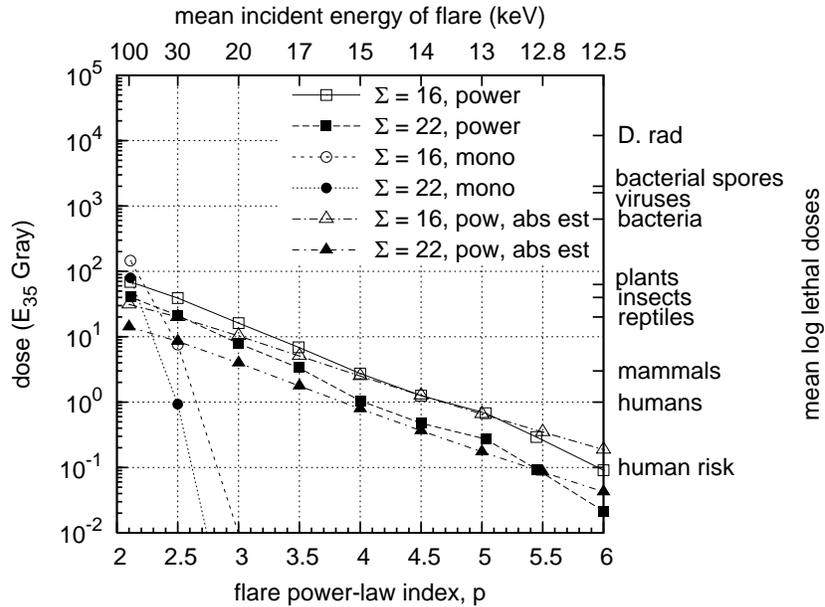}
\caption{Surficial X-ray dose for a flare with 10$^{35}$ erg
total energy release versus spectral index.  (Doses scale linearly
with energy release.)  Two column densities are used: 16 and 22 g
cm$^{-2}$.  Power-law flare spectra are denoted ``power'' in the legend.
Monoenergetic incoming photons with energies equal to the mean incident
flare energy are shown as ``mono.''  This case demonstrates the inaccuracy
of ignoring the full radiative transfer, since the curve decreases
\emph{much} more rapidly than the power-law cases. The final pair of
curves (denoted ``abs est'') uses a power-law incident spectrum with the
radiative transfer done using the analytic, absorption approximation
discussed in \S\ref{sec:doses}. The general trend among all curves is
that the dose decreases with increasing spectral index.  This is because
(as can be seen on the upper horizontal axis) the average energy of the
flare spectrum decreases with increasing spectral index.  The atmosphere
is vastly more opaque to lower-energy X-rays, which dramatically affects
the surficial dose.  The right-hand vertical axis shows tick marks at
the mean-log lethal dose of a few representative terrestrial organisms
(see Table 1).  }
  \label{fig:doses}
\end{figure}

The calculated X-ray doses in our situation are shown in
Fig. \ref{fig:doses} as a function of power-law index for two column
densities: $\Sigma=16$ g cm$^{-2}$ and $\Sigma=22$ g cm$^{-2}$.
The calculation assumes a solar flare with a total X-ray release of
$10^{35}$ erg (see \S\ref{sec:flareprops}).  Again, for flares
of different energy release, the dose simply scales proportionally.
The doses for the shallowest spectra ($2\le p\le 3$) are about two
orders of magnitude higher than those for the steepest spectra ($5\le
p\le 6$), showing the tremendous sensitivity of the dose on the form
of the incident spectrum.  For comparison to the power-law spectra,
we also show monoenergetic spectra with energies equal to the average
energy in the power-law spectra for each spectral index.  The dramatic
decline in the dose with energy for the monoenergetic cases illustrates
the importance of the high-energy tail in the power-law spectra and the
inaccuracy of doing this calculation using a monoenergetic approximation.

The results shown in Fig. \ref{fig:doses} can be confirmed by considering
a 10$^{35}$-erg flare incident on the top of the atmosphere and subject to
only absorption due to Compton scattering and photoabsorption.  Strictly
speaking, the Compton process is a scattering process, not absorption,
but we will neglect photons that backscatter in the atmosphere and
eventually reach the surface in order to get this estimate.  This enables
us to write an integral that approximates the total dose $D$ in terms of
the atmospheric optical depth $\tau_\mathrm{a}(E)$ and the water opacity
$\kappa_\mathrm{w}(E)$: \beq D = \int_{E_\mathrm{min}}^{E_\mathrm{max}}
\kappa_\mathrm{w}(E)\; \exp[-\tau_\mathrm{a}(E)]\; E\; {dN\over dE}\;
dE,\label{eq:dose_approx} \eeq where $N(E)$ is the incident number
spectrum at the top of the atmosphere and $E_\mathrm{min}$ is the
minimum energy considered in the Monte Carlo code.  This integral can
be evaluated numerically to confirm the results given in our Monte Carlo
calculation; we have done this and present it as the final set of curves
in Fig. \ref{fig:doses} marked ``abs est.'' The integral is very close
to the full calculation at all energies, being only different by at most
a factor of two, which suggests that backscattered radiation is only
marginally important to the surficial doses. For the flattest spectra
($p$ near 2), the code gives a higher (and more accurate) dose because it
includes backscattered radiation.  For steeper spectra, the difference
between the above approximation and the code stems from the discrete
number of bins in the surficial spectra in the code and the fact that the
mean incident spectrum energy is very close to the lower cutoff energy,
causing discretization errors in the output spectra when binning photon
energies close to the cutoff.  The integral above, being continuous,
suffers no such errors, so is probably more accurate in this regime,
where absorption dominates anyway.

The horizontal ticks along the right side of Fig. \ref{fig:doses} show
typical mean-log acute lethal doses for a variety of organisms. The
most comprehensive compilation we know of, Sparrow et al. (1967), gives
$D_{90}$ for X-ray and $\gamma$-ray exposure of 79 organisms. We have
expanded and supplemented this with more recent work, especially for
microorganisms, including the review of virus lethal doses by Rohwer
(1984), lethal doses in bacterial spores summarized by Russell (1982),
studies of food pathogenic bacteria by Dion et al. (1994) and Farkas
(1998), studies of Archaean hyperthermophiles by Kopylov et al. (1993)
and DiRuggiero et al. (1997), mammalian studies presented in Fielden and
O'Neill (1991), studies of mutations in human survivors of Hiroshima
and Nagasaki (see Turner, 1995, p. 398), the Chernobyl accident and
Kazakhstan nuclear weapon tests (Dubrova, 2003), and inhabitants of
the Kerala (India) radioactive hotspot (Forster et al., 2002).  We find
these results to be consistent with each other in cases we have checked.

\begin{center}
\begin{table}
\centering
\caption{Acute lethal dose ranges for representative organisms}
\begin{tabular}{ll}
\hline
Organism & Acute Lethal Dose (Gy) \\
\hline
mammals & 1--10 \\
higher plants &  5--700 \\
insects & 1--2000 \\
bacteria, protozoa, algae & 30--15,000 \\
viruses & 150--20,000 \\
bacterial spores & 1,000--4,000\\
\hline
\end{tabular}
\end{table}
\end{center}

Because of the large range of lethal doses found in nature we have chosen
to represent various organisms in Fig. \ref{fig:doses} by the average of
the base-10 logarithm of the lower and upper limits found for that class.
This is an appropriate choice because the available data from Sparrow et
al. (1967) within each taxonomic class are distributed roughly uniformly 
in log lethal dose.  Thus for mammals, with a range of 1--10 Gy, the
number indicated on the log scale is 0.5.   Table 1 lists the approximate
lethal dose ranges for these organisms.  Organisms with extremely large
lethal doses are interesting for long-term survival prospects---e.g. adult
Drosophilia at 1000 Gy, some Archaean hyperthermophiles at 2000--6000 Gy,
\emph{Rubrobacter radiotolerans} at 8000 Gy (Ferriera et al., 1999),
\emph{Deinococcus radiodurans} at 10,000 Gy (depending on cell phase;
see Battista, 1997), and the virus-like proteinaceous infectious particle
(prion) associated with scrapie at 20,000 Gy (Rohwer, 1984).  

Even very low doses can be important.  Doses for mutation and clustered
DNA damage can be much smaller than the lethal dose. For example, in
humans, the whole-body acute lethal dose is about 1--2 Gy, but mutations
(Sparrow et al., 1972; Sankaranarayanan, 1982; Forster et al., 2002),
chromosomal abnormalities in blood lymphocytes (Violot et al., 2005),
and clustered DNA damage (Sutherland et al., 2000) occur at only around
0.01--0.2 Gy, and germ-line mutation doubling is believed to occur
around 0.5 Gy (UNSCEAR 2001; Sankaranarayan and Chakraborty, 2000; but
see Forster et al., 2002).  So doses even smaller than the ``humans''
level in Fig. \ref{fig:doses} should be considered hazardous for human
exploration of Mars.  We indicate this level at 0.1 Gy with the label
``human risk'' and emphasize that this value is illustrative, not
definitive.

We have ignored the dependence of lethal dose on dose rate.  Sparrow
et al. (1967) claim no evidence for a significant dependence in the
literature they compiled, but it is undoubtedly true that at some level
a dose rate dependence occurs, as has been observed in mammalian and
other organisms.  And since the duration of solar flares is random, it is
difficult to convert received fluences into fluxes, which are needed for
dose rate calculations, hence we have ignored this and focused entirely
on time-integrated quantities in this work.

Finally, recall the possible correlation between hard X-ray flux and
spectral index found in the RHESSI data by Battaglia et al. (2005)
mentioned in \S\ref{sec:fp_spectra}.  If flares with a larger total energy
release do indeed typically have flatter spectra, then they would have
higher mean photon energies.  The Martian atmosphere is more transparent
to X-rays of higher energies, so this effect would boost the surficial
doses received from larger flares.  So in this sense our results are an
upper limit to the flare releases required to produce given doses.

\subsection{Mean Times Between Lethal Events}\label{sec:lethal_wait}

The calculated surficial doses are very sensitive to the flare spectral
index $p$, which varies greatly and unsystematically with other flare
parameters, so a better way to determine the lethality risk is to
integrate the frequency of events delivering a given dose for each $p$
over the probability distribution of $p$ values.  We take this probability
distribution to be the uniform distribution, which is roughly consistent
with results shown in Lee et al. (1993), Bromund et al. (1995), Veronig
et al. (2002), and Qiu et al. (2004) and other references given there.
This procedure yields the total frequency of lethal events for an organism
at the Martian surface due to all flares.

To estimate flare timescales as a function of release $W$, we normalize
to one 10$^{32}$-erg solar flare per decade.  Extrapolating the average
HXR frequency-energy statistic $dN/dW$ (see \S\ref{sec:flareprops})
gives the average time between events of energy release $W_{32}$
(in units of 10$^{32}$ erg) as $T(W) \simeq 10\,W_{32}^q$ yr, with $q$
between about 1.6 and 1.8, and the mean time between events at least
as large as $W_{32}$ as \beq T(\ge W_{32}) = 10\, (q-1)\,W_{32}^{q-1}
\mathrm{yr}.\label{eq:meantime}\eeq This agrees with the estimate
of Hudson (1991) and is broadly consistent with the dozen or so
10$^{31}$--10$^{32}$ erg events that have been observed since GOES soft
X-ray monitoring began in 1976.  Upper limits on proton fluences in
lunar rocks, tree ring records of $^{14}$C, and the requirement of finite
energy suggest that the frequency-energy release relation $dN/dW$ must
steepen above some energy (Reedy et al., 1983; Hudson, 1991; Aschwanden,
1999), but the value of that energy is unknown.

\begin{figure}
  \centering
  \includegraphics[height=0.6\textheight,angle=270]{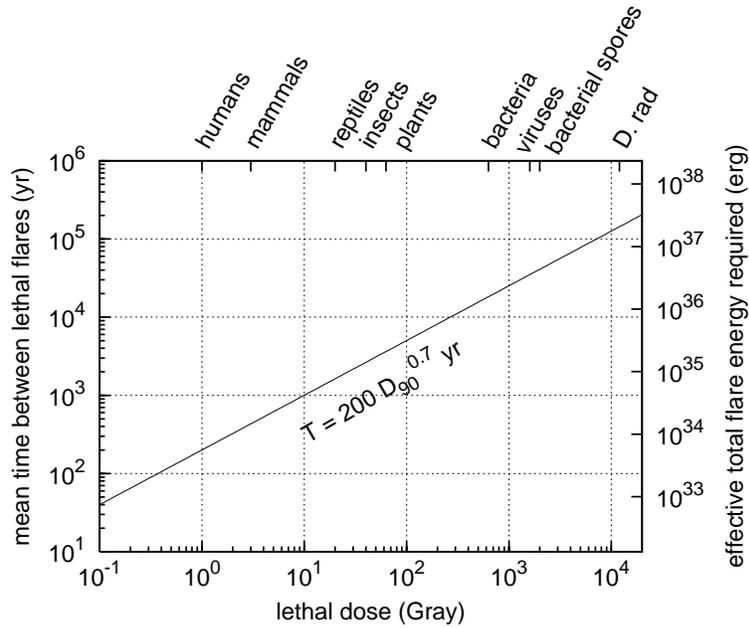}
\caption{Mean time between flare events that deliver a given dose.
The dose delivered depends on the spectral index, such that flares with
larger spectral index must have a larger total energy to deliver a given
dose (see Fig. \ref{fig:doses}).  We have integrated over the distribution
of spectral indices, and accounted for the dose dependence on this index,
in order to find a mean time between events that deliver a given dose
without specifying the spectral index (see \S\ref{sec:lethal_wait}).  The upper
horizontal axis shows the lethal doses of a variety of organisms for
reference (see Table 1). As expected, the time between events increases
as the dose necessary for lethality increases.}
  \label{fig:mean_lethal_wait}
\end{figure}

Figure \ref{fig:mean_lethal_wait} graphically shows the mean time between
lethal events for the same set of organisms given in Fig. \ref{fig:doses}
as a function of surficial dose, for an atmospheric column density
$\Sigma$ of 16 g cm$^{-2}$.  To produce this plot, we use the mean
of the empirical values of frequency-energy release slope, so $q=1.7$
in Eq. \ref{eq:meantime}.  The right-hand side vertical axis shows
the corresponding flare energy releases $W$ required for lethality.
Again, this result is integrated over the spectral index distribution.
Additionally, for a column density of 22 g cm$^{-2}$, the attenuation is
roughly a factor of 2--3 higher (as can be see in Fig. \ref{fig:doses}),
so the required lethal dose would be roughly a factor of 2--3 higher,
implying a mean waiting time that is $2^{0.7}$--$3^{0.7} =$ 1.6--2.2
times longer.

The mean time to lethality is strongly weighted toward the flattest ($p
\sim 2$) flares.  The atmosphere is more transparent to the flatter flares
(because they have a higher average energy), so a lethal fluence at the
surface requires less total energy release at the sun. Since flares with
lower total energy release are much more common, chances are that the flare
that finally delivers the lethal dose will have a smaller spectral index.
We show this explicitly below.

To derive the equation for the mean time between flares of a given dose,
as shown in Fig. \ref{fig:mean_lethal_wait}, we first convert the waiting
time distribution given in Eq. \ref{eq:meantime} back to the empirical
flare energy release-mean frequency relation: \beq \overline{\nu}(W) =
k_W W_{32}^{-q}\; \mathrm{yr}^{-1}, \eeq where once again we use $q=1.7$
and we estimate $k_W\sim 0.1$.  The flares are roughly distributed
uniformly by spectral index $p$, and since $p$ ranges from $2$ to $6$,
we can write the probability distribution function of $p$ as $f(p) = 1/4$.
The frequency distribution for the flares as a function of energy $W$ and
spectral index $p$ is then \beq \overline{\nu}(W,p) = \overline{\nu}(W)\,
f(p),\eeq where we have assumed that $W$ and $p$ are uncorrelated.

To predict how often solar X-ray flares can deliver a particular lethal
dose, we must transform this distribution to a function of dose $D$.
This requires knowledge of the relation between total flare energy
release and surficial dose for each spectral index $p$.  In our case,
this relation comes from our radiation transport code. An empirical
fit to our calculations for a column density of 16 g cm$^{-2}$ shows that the
dose in Gray as a function of flare energy is \beq D(W,p) = {W_{32}\over
k_D}\; e^{-p/p_0}\; \mathrm{~Gy},\eeq with $k_D=0.23$ and $p_0=0.53$. We
can write the energy required to deliver a given dose by inverting
this equation.  Then, using the standard technique for transforming
probability distribution functions, we have \beqa \overline{\nu}(D,p) &=&
\overline{\nu}[W(D,p),p] \left|\frac{\partial(W,p)}{\partial(D,p)}\right|
\\ &=& \overline{\nu}[W(D,p)] \left|\frac{dW}{dD}\right| f(p)\\ &=&
\frac{k_W k_D^{1-q}}{4} D^{-q} \exp[(1-q)p/p_0]\; \mathrm{~yr}^{-1}.\eeqa

Next, we would like to know how many flares
of all spectral indices occur each year, but now the dose and spectral
index are not independent, so we must integrate, instead of just scaling
by $1/4$: \beqa \overline{\nu}(D) &=& \int_{2}^{6} \overline{\nu}(D,p)\;
dp \\ &=& {k_W k_D^{1-q} p_0 \over 4(q-1)} D^{-q} \left\{\exp[2(1-q)/p_0]
- \exp[6(1-q)/p_0]\right\}\; \mathrm{yr}^{-1}.\eeqa

All flares that deliver doses above the lethal dose are most definitely
lethal, so we must integrate the flare dose-mean frequency distribution
above the lethal dose (we use $D_{90}$ here) to find the total number
of flares that occur on average per year that deliver at least the
lethal dose: \beqa \overline{\nu}(\ge D_{90}) &=& \int_{D_{90}}^\infty
\overline{\nu}(D)\; dD \\ &=& {k_W k_D^{1-q} p_0 \over 4(q-1)^2}
\left\{\exp[2(1-q)/p_0] - \exp[6(1-q)/p_0]\right\}\; D_{90}^{1-q} \;
\mathrm{~yr}^{-1}.\eeqa Finally, to get the mean time between lethal
events, simply take the inverse of this frequency: $\overline{T}(\ge
D_{90}) = 1/\overline{\nu}(\ge D_{90})$. For the case discussed in this
work, $k_W\sim 0.1$ and $q=1.7$.  Thus the typical time between lethal
flares as a function of lethal dose in Gray is \beq \overline{T}(\ge
D_{90}) \simeq 200\; D_{90}^{0.7}\; \mathrm{~yr}.\eeq  This time is
much shorter than an estimate using just a particular $p$ value yields.
For example, using $p=4$ instead of integrating over the distribution
of $p$ would give about 3000 yr for the coefficient above instead of
200 yr.  Properly treating the spectral indices as a distribution in
the calculation is important, rather than just assuming that $p=4$ is
the representative index and making simple estimates based on that.
In fact, the true waiting times are probably a bit shorter than our
estimate, since the distribution of spectral indices isn't exactly flat,
but rather is peaked slightly at lower indices.  In other words, the
average flare spectrum may be somewhat harder than we take here.

Figure \ref{fig:mean_lethal_wait} indicates that a solar X-ray flare
of 10$^{38}$ erg would be needed to kill any organism as resistant
as \emph{D. rad}, bacterial spores, and the most resistant viruses
(see Table 1) on the surface of Mars.  Such flares would occur at least
every  $\sim$ 200 kyr if our extrapolation of the solar frequency-energy
release scaling could be extended to these extreme energies.  Since such
energies have not been observed in the sun or other (non-binary) stars,
we conclude that such extremely radiation-resistant microbial life on
Mars may not be affected by solar flares.  Bacteria and protists with the
adopted mean log lethal dose require slightly less extreme flare energies
and have corresponding times to lethality of about 20 kyr.  Reptiles,
insects, and plants require more reasonable flare energies, especially
considering that many have lethal doses much less than the mean log (see
Table 1 and Sparrow et al.,  1967).  Flares of only 10$^{34}$--10$^{35}$
erg will result in mutations and lethality for mammals, reptiles, insects,
and plants, and may occur every few hundred years.  Similarly, a 10$^{34}$-erg flare, which can be lethal to humans, might occur every 200 yr on
average.  We conclude that no organisms other than the most resistant
bacteria and viruses could have survived on the surface of Mars during the
past $\sim 10^5$ yr, which is almost certainly much smaller than the time
during which Mars has had its current low-column-density atmosphere.

Additionally, a correlation between hard X-ray flux and total flare
energy release in which larger flares have harder (flatter) spectra
(e.g., Battaglia et al., 2005) would decrease the time between lethal
events. Larger flares would be more penetrating and would deliver higher
doses than we have estimated, which would decrease the energy release
needed for a given dose. Since less energetic flares are more common,
lethality would occur more often.

\subsection{Shielding and Dose Rate Considerations}

\subsubsection{Soil and CO${_2}$ Ice Shielding}

So far, we have ignored possible protection strategies, such as
radiation shielding sources.  As an illustrative example, we take
the Mars Pathfinder rover soil sample data of W\"anke et al. (2001)
and calculate the opacity of this material to incoming flare spectra.
Based on the composition of the soils in Table I of W\"anke et al. (2001),
we find that the samples have an effective atomic number of 15.2
(the $Z$ that, when used in Eq. \ref{eq:sigma_pa}, yields the mean
cross section of the soil obtained by element-by-element addition).
The combined photoabsorption and Compton scattering opacity of this
material can be calculated.  We use the analytical approximation in
Eq. \ref{eq:dose_approx} to calculate the transfer of the surficial
radiation through the soil. Using an atmospheric column density of 16 g
cm$^{-2}$, we find $1/e$ (i.e., optical depth unity) shielding columns
for X-ray flares of spectral index $p=2$ and $p=6$ to be 7.7 g cm$^{-2}$
and 3.6 g cm$^{-2}$, respectively.  For an atmospheric column density
of 22 g cm$^{-2}$, the same calculation gives only a slightly different
result: 8.1 g cm$^{-2}$ and 3.9 g cm$^{-2}$, respectively. With a typical
density of terrestrial andesites of 2.7 g cm$^{-3}$, these soil columns
suggest that around a few cm of regolith is enough to provide $1/e$
attenuation and that tens of cm will reduce the dose by 99\% or more.
For comparison, the possible ice layers found by Boynton et al. (2002)
at 40--150 g cm$^{-2}$ below the surface should be very well shielded.

CO$_2$ ice has an effective $Z$ of 7.4, roughly half that of the soil,
so $1/e$ shielding requires a slightly larger ice column.  For an
atmospheric column density of 16 g cm$^{-2}$, we find $1/e$ shielding
columns of CO$_2$ ice for spectral indices $p=2$ and $p=6$ to be 8.2 g
cm$^{-2}$ and 4.6 g cm$^{-2}$, respectively.  For a column density of 22
g cm$^{-2}$, we get 8.6 g cm$^{-2}$ and 4.9 g cm$^{-2}$, respectively.
Roughly three times the thickness of ice would be needed for the same
$1/e$ shielding as corresponding soil columns due to the different
densities.  The difference is smaller than expected from the $Z^3$
dependence of the photoabsorption cross section because a large fraction
of photons in the surficial spectrum, even for steep flare spectra,
has energies high enough that the radiative transfer is dominated by
Compton scattering, a process insensitive to elemental composition.

Altitude can change the atmospheric column density and have a measurable
effect on the dose.  The column density above Hellas Basin, the lowest
topographical point on Mars, is roughly 8 to 14 g cm$^{-2}$ larger
(depending on season) than the average and consequently reduces the dose
to a factor of about $1/7$ to $1/3$ of the dose at the mean surface level.
This suggests that organisms in low areas would be better protected,
but not completely safe. Hellas Basin is an extreme example, and more
generally we expect topographic altitude variations to give less
than a factor of two variation in dose for a given flare.

\subsubsection{Manned Mission Shielding}

Astronaut protection is a critical issue for long-term, manned
space missions.  The total mass of a spacesuit depends mostly on the
column density of the component material, so, given a fixed shielding
requirement, the suit's mass will be minimized by maximizing the suit's
opacity to incoming radiation.  Typical next generation spacesuit designs
provide only up to a few g cm$^{-2}$ of mostly low-$Z$ shielding material
(see Ross et al., 1997, for a review).  The lightweight and mostly low-$Z$
suits are designed to provide efficient protection from incoming energetic
particle radiation (such as cosmic rays).   Charged particles lose more
energy per unit column density traversed in materials with lower $Z$
because the energy loss is roughly proportional to $Z/A$, where $A$ is
the mean atomic mass, and lighter elements maximize $Z/A$.  Additionally,
low-$Z$ materials produce less secondary radiation (mostly neutrons) and
lower absorbed doses when incident particles cause nuclear reactions in
the shielding material (see, e.g., Schimmerling et al., 1996, and Wilson
et al., 2001).

X-ray radiation is most efficiently stopped by high-$Z$ materials.
Neglecting engineering considerations and particle radiation, the
ideal shielding material for X-rays would maximize the quantity
$Z^3/A$, because the photoabsorption opacity, which dominates at
these energies, has roughly this dependence on $Z$ and $A$, as can be
seen in Eq. \ref{eq:sigma_pa}.  Thus, heavier elements are favored for
X-ray shielding, since atomic mass $A$ increases much slower than $Z^3$
across the periodic table.  Spacecraft and surface structures, such as
``Marsbases,'' would probably contain a significant amount of high-$Z$
material, and thus could provide adequate flare X-ray shielding if
surrounding structures provide a shielding column of greater than tens of
g cm$^{-2}$ of $Z \gtrsim 10$ material.  For example, polymers, water, and
lower-$Z$ metals, such as magnesium and aluminum, would be less desirable
than titanium or steel alloys (especially those high in molybdenum).
And even a few g cm$^{-2}$ of high-$Z$ material is inadequate should
a large solar X-ray flare occur during a spacewalk or ``Marswalk.''
Balancing the two contrasting shielding requirements for particles and
photons while minimizing the weight and rigidity of spacesuits may prove
to be difficult.

All of these considerations also apply to moon missions and space missions
elsewhere in the solar system, with doses being even higher outside of
Mars' protective atmosphere and inversely proportional to the square of
the distance to the sun.

\subsubsection{X-ray Flares and Cosmic Rays: Dose Rate Differences}

The mean times to lethality we give refer only to the surface of Mars on
the hemisphere facing the sun for organisms with short generation times.
Lethality from a solar flare contrasts with the situation for cosmic-ray
exposure; the difference is one of acute versus extended weak exposure.
Flares are intermittent bursts, and lethality will occur during the flare
if the flare is sufficiently energetic. The organism lifetime does not
affect lethality unless it is less than the mean flare duration (about 10--60 min for the flares considered here), and even most bacteria have
longer generation times.  For example, the average generation time for
many \emph{E.~coli} strains is about an hour.  Cosmic rays, however,
cause damage over long periods of time from constant, relatively
small, surficial fluxes.  Organisms might be able to continuously
repair the cosmic-ray damage (as on the present Earth), while the high
dose rates of flare exposure may overwhelm genomic repair systems.
The Saganti et al. (2004) and Pavlov et al. (2002) calculations cited
in \S\ref{sec:intro} give Galactic cosmic-ray dose rates of 0.06--0.2 Gy
yr$^{-1}$.  Taking the representative dose rate to be 0.1 Gy yr$^{-1}$,
the time to accumulate a lethal dose from cosmic rays is longer than
the lifetime of all but the most sensitive mammals and long-lived
plants.  

Pavlov et al. (2002) also suggest that living organisms would survive
cosmic-ray irradiation because of their short lifetimes, and only dormant
organisms, such as spores, with their inactive genomic repair systems,
would be killed by the slowly accumulating damage.  In contrast, solar
flares with very high energies have dose rates high enough to kill living
organisms (because the dose is so acute) and possibly contribute to or
even dominate the death of dormant organisms. Dormant organisms
have radiation tolerances generally much larger than active organisms,
but they accumulate damage over a much longer time period. In this case
the total dose would be due to a steady dose rate of cosmic rays plus
the accumulated acute doses of a large number of individual X-ray
flares over the period of dormancy.

\section{Summary and Implications}

Continuing astronomical observational programs are recording larger
and larger X-ray flares from stars of solar mass and smaller.  The most
energetic flares observed on the sun are almost large enough to cause
mutation and lethality in higher organisms on the surface of Mars, and
extrapolation to higher energies implies greater, but less frequent,
biological damage.

Using a previously developed Monte Carlo radiative transport code
(Smith et al., 2004b), we have calculated the surficial X-ray spectra
and corresponding biological doses due to solar flares with a range of
spectral indices. We compared our results to the average lethal doses
for a representative range of terrestrial organisms.  We find that,
if the sun has flares much larger in energy output than those observed
(as supported by observations of other stars), organisms with a range of
radiological tolerances, from mammals to plants, insects, and reptiles
to even high-tolerance bacteria, spores, and viruses could be killed
on the hemisphere of Mars facing the sun during the flare.  Though a single
flare leads to only partial irradiation, flares are often clustered in
time, so it may be possible to irradiate the whole surface over the course of
weeks during a period of extreme solar activity.

Estimates were given for the mean time between lethal events based
on extrapolating the observed recurrence frequency-energy release
distribution for solar flares.  Precise results are difficult because
of the random nature of flare occurrence and spectral indices, so we
provide mean times between lethal flare events integrated over the
spectral index distribution.  In addition, the estimated mean times
between flares of such energies are orders of magnitude less than the age
of the solar system, so may have occurred 10$^3$--10$^7$ times for the
period during which Mars has had its present thin atmosphere, depending
on the maximum possible solar flare energy and required lethal dose.

Finally, we showed that a column thickness of $\sim$ 4--8 g cm$^{-2}$
of Martian soil would be sufficient to provide $1/e$ attenuation of
the surficial flare energy.  The conflicting shielding requirements for
particle radiation and the X-rays discussed here leads to a difficultly
designing safe spacesuits for manned Mars or lunar exploration, but
designing safe ``Marsbases'' and other ground structures should be easier.

\section*{Acknowledgements}

DSS was supported by the NSF Graduate Student Research Fellowship and
Harrington Doctoral Fellowship Programs. JMS was supported by the NASA
Exobiology Program, Grant NNG04GK43G.  This work was carried out as part
of the research of the NASA Astrobiology Institute Virtual Planetary
Laboratory Lead Team, which is supported through the NASA Astrobiology
Institute.

\section*{References}

\parindent=0in
\newcommand{\apj}{Astrophys.~J.}
\newcommand{\aj}{Astr.~J.}
\newcommand{\apjl}{Astrophys.~J.~Lett.}
\newcommand{\apjs}{Astrophys.~J.~Supp.}
\newcommand{\icar}{Icarus}
\newcommand{\sci}{Science}
\newcommand{\nat}{Nature}
\newcommand{\newa}{New Astr.}
\newcommand{\mnras}{Mon.~Not.~Roy.~Astr.~Soc.}
\newcommand{\araa}{Ann.~Rev.~Astr.~Astrophys.}
\newcommand{\aanda}{Astr.~Astrophys.}
\newcommand{\aandasupp}{Astr.~Astrophys.~Supp.}
\newcommand{\jgr}{J.~Geophys.~Res.}
\newcommand{\pasp}{Pub.~Astr.~Soc.~Pacific}
\newcommand{\oleb}{Origins Life Evol.~Biosphere}
\newcommand{\pss}{Plan.~Sp.~Sci.}

Aschwanden, M.J., 1999. Nonthermal flare emission. In: Strong, K.T., Saba, J.L.R.,
Haisch, B.M., Schmelz, J.T. (Eds.), The Many Faces of the Sun.
Springer, NY. 273--300.

Aschwanden, M.J., Tarbell, T.D., Nightingale, R.W., and 5
others, 2000. Time variability of the ``quiet'' Sun observed with
TRACE. II.~Physical parameters, temperature evolution, and energetics
of extreme-ultraviolet nanoflares. \apj. 535, 1047--1065.

Audard, M., G\"udel, M., Drake, J.J., Kashyap, V.L. 2000.
Extreme-ultraviolet flare activity in late-type stars. \apj. 
541, 396--409.

Baker, D.N., Kanekal, S.G., Li, X., Monk, S.P., Goldstein, J., Burch,
J.L., 2004. An extreme distortion of the Van Allen belt arising from the
``Halloween'' solar storm in 2003.  Nature. 432, 878--881.

Battaglia, M., Grigis, P.C., Benz, A.O., 2005. Size
dependence of solar X-ray flare properties. \aanda, in press
(http://arxiv.org/astro-ph/0505154).

Battista, J.R., 1997. Against all odds: the survival strategies of
\emph{Deinococcus radiodurans}. Ann.~Rev.~Microbiology. 51, 203--224.

Becker, D., Sevilla, M.D., 1993. The chemical consequences of radiation
damage to DNA.  Adv.~Radiation Biol. 17, 121--180.

Boynton, W.V., Feldman, W.C., Squyres, S.W., and 22 others, 2002.
Distribution of hydrogen in the near surface of Mars: Evidence for
subsurface ice deposits. Science. 297, 81--85.

Bromund, K.R., McTiernan, J.M., and Kane, S.R., 1995. Statistical
studies of ISEE 3/ICE observations of impulsive hard x-ray solar
flares.  \apj. 455, 733--745.

Christian, D. J., Mathioudakis, M., Jevremovic, Dupuis, J., Vennes, S.,
Kawka, A., 2003. The extreme-ultraviolet continuum of a strong stellar
flare. \apjl. 593, L105--L108.

Cockell, C.S., Catling, D.C., Davis, W.L., Snook, K., Kepner, R.L.,
Lee, P., and McKay, C.P., 2000. The ultraviolet environment of Mars:
Biological implications past, present, and future.  Icarus. 146,
343--359.

Cordoba-Jabonero, C., Lara, L.M., Mancho, A.M., M\'arquez, A.,
Rodrigo, R., 2003. Solar ultraviolet transfer in the Martian
atmosphere: Biological and geological implications. \pss. 51,
399--410.

Cravens, T.E., Maurelis, A.N., 2001. X-ray emission from scattering and
fluorescence of solar x-rays at Venus and Mars.  Geophys. Res. Lett.
28, 3043--3046.

Crosby, N.B., Siegmund, O.H.W., Vedder, P.W., Vallerga, J.V.,
1993. Extreme Ultraviolet Explorer deep survey observations of a large
flare on AU Microscopii. \apjl. 414, L49--L52.

De Angelis, G., Clowdsley, M.S., Singleterry, R.C., Wilson, J.W.,
2004. Mars radiation environment model with visualization.  Adv. Space
Res. 34, 1328--1332.

Dennerl, K., 2002. Discovery of X-rays from Mars with Chandra. \aanda.
394, 1119--1128.

Dion, P., Charbonneau, R., and Thibault, C., 1994.  Effect of ionizing
dose rate on the radioresistance of some food pathogenic bacteria.
Can.~J.~Microbiol. 40, 369--374.  

DiRuggiero, J., Santangelo, N., Nackerdien, Z., Ravel, J., and Robb, F.T.,
1997. Repair of extensive ionizing-radiation DNA damage at
95 C in the hyperthermophilic archaeon \emph{Pyrococcus furiosus}.
J.~Bacteriology. 179, 4643--4645.

Dubrova, Y.E., 2003. Long-term genetic effects of radiation exposure.
Mutation Res. 544, 433--439.

Farkas, J., 1998. Irradiation as a method for decontaminating food.
Int.~J.~Food Microbiology. 44, 189--204.

Favata, F., Reale, E., Micela, G., Sciortino, S., Maggio, A., and
Matsumoto, H., 2000. An extreme X-ray flare observed on EV Lac by ASCA
in July 1998. \aanda. 353, 987--997.

Ferreira, A.C., Nobre, M.F., Moore, E., Rainey, F.A., Battista, J.R.,
da Costa, M.S., 1999.  Characterization and radiation resistance of
new isolates of \emph{Rubrobacter radiotolerans} and \emph{Rubrobacter
xylanophilus}.  Extremophiles. 3, 235--238.

Fielden, E.M., and O'Neill, P., 1991. The Early Effects of Radiation on
DNA. Springer, Berlin.

Forster, L., Forster, P., Lutz-Bonengel, S., Willkomm, H., and Brinkmann,
B.,  2002.  Natural radioactivity and human mitochondrial DNA mutations.
PNAS 99, 13950--13954.

Friedberg, E.C., 2003. DNA damage and repair. \nat. 421, 436--440.

G\"udel, M., Audard, M., Kashyap, V.L., Drake, J.J., and Guinan,
E.F., 2003. Are coronae of magnetically active stars heated by
flares? II.~Extreme ultraviolet and X-ray flare statistics and the
differential emission measure distribution.  \apj. 582, 423--442.

Gunell, H., Holstrom, M., Kallio, E., Janhunen, P., Dennerl, K., 2004. X
rays from solar wind charge exchange at Mars: A comparison of simulations
and observations.  Geophys.~Res.~Lett. 31, L22801--L22804.

Haisch, B., Strong, K.T., and Rodono, M.A., 1991.  Flares on the sun
and other stars.  \araa. 29, 275--324.

Hawley, S.L. and Pettersen, B.R., 1991. The great flare of 1985 April
12 on AD Leonis.  \apj. 378, 725--741.

Hudson, H.S., 1991. Solar flares, microflares, nanoflares, and coronal
heating. Solar Phys. 133, 357--369.

Kanbach, G., Bertsch, D.L., Fichtel, C.E., and 7 others, 1993.
Detection of a long-duration solar gamma-ray flare on June 11, 1991
with EGRET on COMPTON-GRO. \aandasupp. 97, 349--353.

Kopylov, V.M., Bonch-Osmolovskaya, E.A., Svetlichnyi, V.A.,
Miroshnichenko, M.L., and Skobkin, V.S., 1993. Gamma-irradiation
resistance and UV-sensitivity of extremely thermophilic archaebacteria
and eubacteria. Mikrobiologiya. 62, 90--95.

Krucker, S., and Lin, R.P., 2002. Relative timing and spectra of solar
flare hard X-ray sources. Solar Phys. 210, 229--243.

Lee, T.T., Petrosian, V., and McTiernan, J.M., 1993. The distribution of
flare parameters and implications for coronal heating.  \apj.
412, 401--409.

Liebert, J., Kirkpatrick, J.D., Reid, I.N., and Fisher, M.D., 1999. A
2MASS ultracool M dwarf observed in a spectacular flare.  \apj.
519, 345--353.

Lin, R.P., Feffer, P.T., Schwartz, R.A., 2001. Solar hard X-ray bursts
and electron acceleration down to 8 keV. \apjl. 557, L125--L128.

Mancinelli, R.L., and Klovstad, M., 2000. Martian soil and UV
radiation: microbial viability assessment on spacecraft surfaces.
Plan.~Sp.~Sci. 48, 1093--1097.

Mileikowsky, C., Cucinotta, F.A., Wilson, J.W., and 7 others, 2000.
Natural transfer of viable microbes in space. 1.~From Mars to Earth
and Earth to Mars. \icar. 145, 391--427.

Molina-Cuberos, G.J., Stumptner, W., Lammer, H., Komle, N.I., O'Brian,
K., 2001. Cosmic ray and UV radiation
models on the ancient martian surface. \icar. 154, 216--222.

Pagano, I., Ventura, R., Rodono, M., Peres, G., and Micela, G., 1997. A
major optical flare on the recently discovered X-ray active dMe star G
102-21.  \aanda. 318, 467--471.

Pavlov, A.K., Blinov, A.V., Konstantinov, A.N., 2002. Sterilization of
Martian surface by cosmic radiation. \pss. 50, 669--673.

Qiu, J., Liu, C., Gary, D.E., Nita, G.M., Wang, H., 2004. Hard X-ray
and microwave observations of microflares.  \apj. 612, 530--545.

Quillardet, P., Rouffaud, M.A., and Bouige, P., 2003. DNA array
analysis of gene expression in response to UV irradiation in
\emph{Escherichia Coli}. Res. Microbiol. 154, 559--572.

Randall, C.E., Harvey, V.L., and 9 others, 2005. Stratsopheric effects
Of energetic particle precipitation in 2003--2004. Geophys. Res. Lett. 32,
L05802--L05805.

Reedy, R.C., Arnold, J.R., Lal, D., 1983. Cosmic-ray record in solar system
matter. Science. 219, 127--135.


Rohwer, R.G., 1984. Scrapie infectious agent is virus-like in size and
susceptibility to inactivation.  Nature. 308, 658--662. 

Ross, A.J., Webbon, B., Simonsen, L.C., Wilson, J.W., 1997. Shielding
strategies for human space exploration.  Wilson, J.W., Miller, J., Konradi,
A., Cucinotta, F.A. (eds.). NASA CP 3360, 283--296.

Rubenstein, E.P., and Schaefer, B.E., 2000. Are superflares on solar
analogues caused by extrasolar planets? \apj. 529, 1031--1033.

Russell, A.D., 1982. The destruction of bacterial spores. Academic
Press, New York.

Ryan, J.M., 2000. Long-duration solar gamma-ray flares. Sp.~Sci.~Rev.
93, 581--610.

Saganti, P.B., Cucinotta, F.A., Wilson, J.W., Simonsen, L.C., Zeitlin,
C., 2004. Radiation climate map for analyzing risks to astronoauts on
the Mars surface from galactic cosmic rays.  Sp.~Sci.~Rev.  110, 143--156.

Sankaranarayanan, K., 1982. Genetic Effects of Ionizing Radiation in
Multicellular Eukaryotes and the Assessment of Genetic Radiation Hazards
in Man. Elsevier Biomedical, Amsterdam.

Sankaranarayanan, K., Chakraborty, R., 2000. Ionizing radiation and
genetic risks. XI. The doubling dose estimates from the mid-1950s to
present and the conceptual change to the use of human data on spontaneous
mutation rates and mouse data on induced mutation rates for doubling
dose calculations.  Mutat.~Res. 453, 107--127.

Schaefer, B.E., King, J.R., Deliyannis, C.P., 2000. Superflares on
ordinary solar-type stars. \apj. 529, 1026--1030.

Schimmerling, W., Wilson, J.W., Nealy, J.E., Thibeault, S.A., Cucinotta,
F.A., Shinn, J.L., Kim, M., Kiefer, R., 1996.  Shielding against Galactic
cosmic rays.  Adv. Sp. Res. 17, (2)31--(2)36.

Schuerger, A.C., Mancinelli, R.L., Kern, R.G., Rothschild, L.J.,
McKay, C.P., 2003. Survival of endospores of Bacillus subtilis on
spacecraft surfaces under simulated martian environments: Implications
for the forward contamination of Mars. \icar. 165, 253--276.

Setlow, R.B, Pollard, E.C. 1962. Molecular Biophysics.  Addision-Wesley,
Reading, MA.

Smith, D.S., Scalo, J., Wheeler, J.C., 2004a. Importance of
biologically active aurora-like UV emission: Stochastic irradiation of
Earth and Mars by flares and explosions. Origins Life Evol.~Biosphere.
34, 513--532.

Smith, D.S., Scalo, J., Wheeler, J.C., 2004b. Transport of ionizing
radiation in terrestrial-like exoplanet atmospheres. \icar. 171,
229--253.

Sparrow, A.H., Underbrink, A.G., Sparrow, R.C., 1967. Chromosomes and
cellular radiosensitivity. I.~The relationship of D0 to chromosome
volume and complexity in seventy-nine different organisms. Rad.~Res.
31, 915--949.

Sparrow, A.H., Underbrink, A.G., Rossi, H.H., 1972. Mutations induced
in Tradescantia by small doses of X-rays and neutrons: analysis of
dose-response curves.  \sci. 176, 916--918.

Stothers, R., 1980. Giant solar flares in Antarctic ice. Nature. 287,
365--365.

Sutherland, B.M., Bennett, P.V., Sidorkina, O., Laval, J., 2000.
Clustered DNA damages induced in isolated DNA in human cells by low
doses of ionizing radiation.  Proc.~Natl.~Acad.~Sci. 97, 103--108.

Turner, J.E., 1995. Atoms, Radiation, and Radiation Protection. 
Wiley, New York.

UNSCEAR 2001. Hereditary Effects of Radiation.  United Nations,
New York.

Veronig, A., Temmer, M., Hanslmeier, A., Otruba, W., Messerotti, M.,
2002. Temporal aspects and frequency distributions of solar soft X-ray
flares. \aanda. 382, 1010--1080.

Violot, D., M'kacher, R., Adjadj, E., Dossou, J.,
de Vathaire, F., Parmentier, C., 2005. Evidence of increased chromosomal
abnormalities in French Polynesian thyroid cancer patients.  Eur. J.
Nucl. Med. Mol. Imaging. 32, 174--179.

von Sonntag, C., 1987. The chemical basis of radiation biology. 
Taylor and Francis, London.

W\"anke, H., Br\"uckner, J., Dreibus, G., Rieder, R., Ryabchikov, I.,
2001. Chemical composition of rocks and soils at the Pathfinder site. Sp.
Sci. Rev. 96, 317--330.

Ward, J.F., 1999. Ionizing radiation damage to DNA: A challenge to
repair systems.  In: Dizdaroglu, M., Karakaya, A.E. (Eds.), Advances
in DNA damage and repair: Oxygen radical effects, cellular protection,
and biological consequences. Kluwer/Plenum, NY, pp. 431--439.

Wilson, J.W., Cucinotta, F.A., Kim, M.-H.Y., Schimmerling, W., 2001.
Optimized shielding for space radiation protection. Physica Medica. XVII,
67--71.

Woods, T.N., Eparvier, F.G., Fontenla, J., Harder, J., Kopp, G.,
McClintock, W.E., Rottman, G., Smiley, B., and Snow, M.,  2004. Solar
irradiance variability during the October 2003 solar storm period.
Geophys. Res. Lett. 31, L10802--10806.

\end{document}